\documentclass[twocolumn,10pt,prl,aps,showpacs]{revtex4}

\def\3nab{\tilde{\nabla}}

\def\be {\begin{equation}}
\def\ee {\end{equation}}
\def\bea {\begin{eqnarray}}
\def\eea {\end{eqnarray}}

\newcommand{\sfrac}[2]{{\textstyle{#1\over#2}}}
\def\case#1/#2{\textstyle\frac{#1}{#2}}
\usepackage{graphicx}
\usepackage{dcolumn}
\usepackage{bm}
\usepackage{longtable}

\def\cqg{{\em Class. Quantum Grav.\/} }
\def\grg{{\em Gen. Rel. Grav.\/} }
\def\prd{{\em Phys. Rev.\/} {\bf D}}

\def\aph{{\em Ann. Phys. (NY)\/} }
\def\plb{{\em Phys. Lett.\/} {\bf B}}

\begin{document}

\title{A new approach to reconstruction methods in $f(R)$ gravity}

\author{Sante Carloni$^{3}$, Rituparno Goswami$^{1,2}$ and Peter K. S. Dunsby$^{1,2,4}$}

\affiliation{1. Astrophysics, Cosmology and Gravity Centre (ACGC),  
University of Cape Town, Rondebosch, 7701, South Africa}

\affiliation{2. Department of Mathematics and Applied Mathematics,
  University of Cape Town, 7701 Rondebosch, Cape Town, South Africa}

\affiliation{3. Institut d'Estudis Espacials de Catalunya
(IEEC), Campus UAB, Facultat Ci\`encies, Torre C5-Par-2a pl, E-08193 Bellaterra
(Barcelona) Spain}

\affiliation{4. South African Astronomical Observatory,
  Observatory 7925, Cape Town, South Africa.}

\date{\today}

\begin{abstract}
We present a new approach of the reconstruction method based on the use of the cosmic 
parameters instead of a time law for the scale factor. This allows the derivation and 
analysis of a set of new non-trivial cosmological solutions for $f(R)$-gravity. 
A number of simple examples are given.
\end{abstract}
\maketitle
\section{Introduction} 
The $\Lambda$CDM (or {\em Concordance}) Model \cite{concordance} which is based on the Friedmann-Lema\^{\i}tre-Robertson-Walker (FLRW) metric is one of the greatest successes of General Relativity. It reproduces beautifully all the main observational results e.g., the dimming of type Ia Supernovae \cite{sneIa}, Cosmic Microwave Background Radiation (CMBR) anisotropies \cite{cmbr}, Large Scale Structure formation \cite{lss}, baryon oscillations \cite{bo} and weak lensing \cite{wl}). Unfortunately, this model is also affected by significant fine-tuning problems related to the vacuum energy scale and therefore it is important to investigate other viable theoretical schemes compatible with cosmological observations.

A now widely studied alternative to the {\em $\Lambda$CDM} model is based on gravitational actions which are non-linear in the Ricci curvature $R$ and$/$or contain terms involving combinations of derivatives of $R$: the $f(R)$ theories of gravity \cite{DEfR,kerner,teyssandier,magnanoff}.  Such models first became popular in the 1980's because it was shown that they are derived from fundamental physical theories (Like M-theory) and naturally admit a phase of accelerated expansion which could be associated with an early universe inflationary phase \cite{star80}.  The fact that the phenomenology of Dark Energy requires the presence of a similar phase (although only a late time - low energy one) has recently revived interest in these theories. In particular, the idea that Dark Energy may have a geometrical origin, i.e., that there is a connection between Dark Energy and a non-standard behavior of gravitation on cosmological scales has resulted in it becoming a very active area of research  over the past few years (see for example \cite{ccct-ijmpd,review,cct-jcap,otha,perts,cct-mnras}). 

Unfortunately efforts to obtain an understanding of the physics of these theories are hampered by the complexity of the fourth-order field equations, making it difficult to obtain both exact and numerical solutions, which can be compared with observations. Recently, however, progress has been made in resolving this issue using a number of useful techniques. One such method, based on the theory of dynamical systems \cite{DS} has proven to be very successful in providing a simple way of obtaining exact solutions and a (qualitative) description of the global dynamics of these models \cite{Dyn}.

Another interesting technique is the so called {\it reconstruction method} \cite{odintsov}. Here one assumes that the expansion history of the universe is known exactly and one inverts the field equations to deduce what class of $f(R)$ theories give rise to a given FLRW model. The existence of such solutions is particularly relevant because in FLRW backgrounds they typically represent asymptotic or intermediate states in the full phase-space of all possible cosmological evolutions. In \cite{julien} it was found that expansion histories based on a power-law solution for the scale factor and a perfect fluid only exist for $R^n$ gravity, while in \cite{reconstruction2} it was shown that in order to reconstruct a $\Lambda$ CDM expansion history in $f(R)$ gravity, it is necessary to add additional degrees of freedom to the matter sector.

A fundamental limitation of the reconstruction method is that, because of the mathematical steps involved, only very simple cosmic histories (i.e., single power law behaviors) can be successfully connected to a $f(R)$-theory in an exact way \footnote{Naturally one can use numerical methods, but these approaches have their own limitations which will not be considered here.}. On the other hand $f(R)$-theories of gravity, being of order four in general admit  a much richer set of solutions, which have at most four different modes. From this point of view it is clear how reconstructing with a single power law behavior corresponds to a dramatic constraint on the form of the function $f$. As a consequence  one obtains by reconstruction,  Lagrangians which are often of little use due to their complexity. 

In this paper we propose a modification of the classic reconstruction scheme able to generate more general cosmic histories and able to give much simpler results for the form of $f(R)$. This will be done describing these expansion histories by means of the cosmic parameters rather than a function of time. 
\section{Field equations for $f(R)$ FLRW models}
Let us consider the typical action for $f(R)$-gravity in the $c=8\pi G=1$ units:
\begin{equation}\label{lagr f(R)}
\mathcal{A}=\int d^4 x \sqrt{-g}\left[\frac12 f(R)+{\cal L}_{m}\right]\;,
\end{equation}
where $R$ is the Ricci scalar, $f$ is general differentiable 
(at least $C^2$) function of the Ricci scalar and $\mathcal{L}_m$ corresponds to 
the matter Lagrangian. 

In a FLRW universe,  the  field equations take the form
\begin{eqnarray}\label{raych}
&&3\dot{H}+3H^2=-\frac{1}{2f'}\left[\rho+3p+f-f'R+3H f'' \dot{R}\right.\nonumber\\ 
&&~~~~~~~~\left.+3f'''\dot{R}^2+3f''\ddot{R}\right]\label{ray}\;,\\
&&3H^2= \frac{1}{f'}\left[\rho+\frac{Rf'-f}{2}-3H f'' \dot{R}\right]\;,
\end{eqnarray}
i.e., the {\em Raychaudhuri} and {\em Friedmann} equations. Here $H$ is the Hubble parameter, which defines the scale factor $a(t)$ via the standard relation $H=\dot{a}/{a}$, the Ricci scalar is 
\begin{equation}
R=6\dot{H}+12H^2 \label{R}
\end{equation}
and  $f',f''$ and $f'''$  abbreviates $\partial^n f/{(\partial R)^n}$ for $n=1..3$ respectively. The {\em energy conservation equation} for standard matter
\begin{equation}\label{cons:perfect}
\dot{\rho}=-3H\left(\rho+p\right)
\end{equation}
closes the system.

It is interesting to note that the Raychaudhury equation can be obtained by adding the Friedmann equation to its time derivative and using the Energy conservation equation and the definition of the Ricci scalar. Hence, any solution of the Friedmann equation  automatically solves the Raychaudhuri equation. Thus, in the reconstruction process we only need to solve  the Friedmann equation, as that guarantees a solution to the other equations.   

\section{Reconstruction strategies}

In this section we introduce, after reviewing briefly the standard reconstruction  method, a new way of obtaining the form of the function $f(R)$ starting from conditions on the dynamics of the cosmic parameters. We also give some specific examples with each of these strategies, which shows that this can be a powerful tool for generating  exact solutions in fourth order gravity.   

\subsection{The classical reconstruction method}
In the classical reconstruction method \cite{odintsov}, the function $f(R)$ is derived  from a given solution 
$a=a(t)$. An explicit reconstruction is possible if and only if the function $R(t)$, as obtained by substituting the solution in equation (\ref{R}), is invertible analytically. That is, if we have an explicit function $g$ such that $t=g(R)$. When this is possible, we can write all the functions, (namely $a(t)$, $H(t)$, $\dot{R(t)}$) in terms of the variable $R$ and the Friedmann equation becomes 
\bea
&&3H[g(R)] \dot{R}[g(R)]\; f''  +f' \left\{3H[g(R)]^2-\frac{R}{2}\right\} +\frac{f}{2} \nonumber\\
&&~~~~~~~~~=\rho_0g(R)^{-3 (1 + w)}\;,
\eea
where we have assumed the matter to be a perfect fluid with equation of state $p=w\rho$. The above equation is now a second order differential equation for the function $f(R)$, the solution of which gives the class of theories of gravity for which the given function $a=a(t)$ is an exact solution. 

As an example let us consider the power law solution $a(t)=a_0t^m$ ~\cite{julien}, 
From (\ref{R}) we see that the Ricci scalar is invertible and is given by
\begin{equation}
R= 6m(2m-1)t^{-2} \equiv \alpha_m t^{-2}\;.
\label{R_pl}
\end{equation}
Solving the Friedmann equation we obtain the following 
general solution \cite{julien}
\bea
f(R)&=&A_{mw}\left(\frac{R}{\alpha_m}\right)^{\sfrac{3}{2} m (1 + w)} + 
C_1R^{\sfrac{3}{4}-\sfrac{m}{4}+\frac{\sqrt{ \beta_m}}{4}}\nonumber\\
&&+\frac{2}{\sqrt{\beta_m}}C_2R^{\sfrac{3}{4}-\sfrac{m}{4}-\frac{\sqrt{ \beta_m}}{4}}\;,
\label{f(R)}
\eea
where $A_{mw}$ and $\beta_m$ are constants depending on the values of $m$ and $w$ and 
$\rho_0$, and $C_{1,2}$ are arbitrary constants of integration. This is a lucky coincidence. 
Considering more general solutions leads at best to a combination of hypergeometric functions (see e.g. \cite{odintsov}).

\subsection{Reconstruction from the condition $\dot{a}=h(a)$}

Let us consider now consider the relation $\dot{a}=h(a)$ instead of $a=a(t)$. This equation relates 
the Hubble parameter to the scale factor. We can easily calculate the Ricci scalar  in this case as
\be
R(a)=6\left(\frac12\frac{(h^2)_{,a}}{a}+\frac{h^2}{a^2}\right)\;.
\label{R_a}
\ee
Like the previous case, an explicit reconstruction is possible if the function $R(a)$, 
is invertible analytically, that is, if we have an explicit function $g$ such that $a=g(R)$.
However, this inversion is now possible in cases which were forbidden before. 

The Friedmann equation can now be written as
\bea
&&3\frac{h[g(R)]}{g(R)} R_{,a}[g(R)]h[g(R)]f'' + f'\left\{ 3\frac{h[g(R)]^2}{g(R)^2}+\frac{Rf'}{2}\right\}\nonumber\\
&&~~~~~~~~~~ +\frac{f}{2}=\rho_0 \,g(R)^{-3(1 + w)}\;.
\label{Fried_a}
\eea
The solutions to the above equation gives the class of theories of gravity 
for which the condition $\dot{a}=h(a)$ is satisfied. This last relation can be then integrated to obtain the corresponding scale factor :
\be
t=\int_0^a \frac{da}{h(a)}\;.
\ee
As an example let us investigate for which class of model a dust-like matter ($w=0$) behaves in the following way
\be
\dot{a}=\frac{2\Omega}{\sqrt{\Lambda}}\sqrt{a-\Lambda a^2}\;.
\ee
As one can easily see, this condition corresponds to a cyclic universe with the scale factor
\be
a(t)=\frac{1}{\sqrt{\Lambda}}\sin^2(\Omega t)\;.
\ee
Using equation (\ref{R_a}), we get an analytically invertible Ricci scalar
\be 
R(a)=\frac{12\Omega^2(3-4\Lambda a)}{\Lambda a}\;.
\ee
Substituting everything in equation (\ref{Fried_a}), we find the particular solution to be
\be
f(R)=\alpha_1 R+\alpha_2 R^2 + \alpha_3 R^3 +\alpha_4\;,
\ee
where $\alpha_n(n=1..4)$ are constants depending on $\Lambda$, $\Omega$ and $\rho_0$.

\subsection{Reconstruction from the condition $\dot{H}=h(H)$.}

Following the line of reasoning above, one can also imagine describing the scale factor as a differential equation for the parameter $H$ e.g., $\dot{H}=h(H)$. In this case, we can immediately find the Ricci Scalar in terms of the Hubble parameter:
\be
R(H)=6h(H)+12H^2
\ee
and an explicit reconstruction is possible if the above relation is analytically invertible, that is, we have an explicit function $g$ such that $H=g(R)$. The scale factor in terms of the Ricci scalar can then be found solving the integral
\be
a(R)=\exp\left[\int \frac{g(R)dg(R)}{h(g(R))}\right]\;.
\ee
Substituting for the above quantities, the Friedmann equation becomes
\bea
&& 3g(R)  R_{,H}[g(R)]h[g(R)]f''+f'\left\{3g(R)^2-\frac{Rf'}{2}\right\}\nonumber\\
&& ~~~~~~~~~~+\frac{f}{2}=\frac{\rho_0}{a(R)^{3(1+w)}}\;.
\eea
Solving the above equation we obtain the $f(R)$ theory which admits a solution compatible with the condition above. To solve for the scale factor as a function of time we express this condition as a differential equation for a(t):
\be
6\left(\frac{\ddot{a}}{a}-\frac{\dot{a}^2}{a^2}\right)=h(\frac{\dot{a}}{a})\;.
\label{ah}
\ee
Let us illustrate this with a simple example by assuming a vacuum ($\rho=0$) universe and let the 
condition be
\be
\dot{H}=m\;,
\ee
where $m$ is a constant. In this case we get a Ricci scalar which is invertible and the solution 
of the Friedmann equation gives the following theory of gravity:
\be
f(R)=\alpha_1 R+\alpha_2 R^2 + \alpha_3\;,
\ee
where the constants  $\alpha_n(n=1..3)$ depends on $m$.  Now solving for the scale factor 
we obtain
\be
a(t)=a_0\exp\left(\frac{m}{2}(t^2-C_1t)\right)\;,
\ee
which, for $m>0$, represents a universe bouncing in the past. 

Note that this solution, unlike the previous ones contains two integration constants.  This is the consequence of the fact that $\dot{H}=h(H)$ is a second order differential equation in $a$ \footnote{In fact in general one would expect a general solution of fourth order gravity to contain four different modes and integration constants. The reason why we find a solution with fewer modes than four is only due to the fact that our initial assumption automatically selects a solution in which one or more integration constants can be put to zero, without compromising the fulfillment of the cosmological equations.}.

\subsection{Reconstruction from the condition $\dot{q}=h(q)$}

It is now clear how we can further generalize the above strategy. One can now give a condition
on the dynamics of the {\it decelaration parameter} $q=-\ddot{a}a/\dot{a}^2$. 

Supposing $\dot{q}=h(q)$, we can integrate to find the Hubble parameter in terms of $q$ as
\be
\frac{1}{H(q)}=\int\frac{1+q}{h(q)}dq\;.
\ee
The Ricci scalar in terms of $q$ is given by
\be
R(q)=6H(q)^2(1-q)\;.
\label{R_q}
\ee
For explicit reconstruction,  the above equation should be analytically invertible as in other 
cases, so that $q=g(R)$. We can then solve for the scale factor in terms of the Ricci scalar
as
\be 
a(R)=\exp\left(\int\frac{H(g(R))}{h(g(R))}dg(R)\right)\;.
\ee
Substituting into the Friedmann equation, we obtain
\bea
3H[g(R)] R_{,q}[g(R)]h[g(R)]f''&& \nonumber  \\
+f'\left\{3H(g(R))^2-\frac{Rf'}{2}\right\}+\frac{f}{2}&=&\frac{\rho_0}{a(R)^{3(1+w)}}\;,
\eea
the solution of which gives the required theory.

As an illustration let us assume
\be
\dot{q}=m(1+q)\sqrt{q}\;,
\ee
where $m$ is a constant. Reconstructing the theory for a universe filled with 
dust-like matter, we get the following function $f(R)$:
\be
f(R)=\alpha R\sqrt{R-3m^2}\;,
\ee
where $\alpha$ is a constant depending on $\rho_0$. Using equation (\ref{R_q}), we can then 
solve for the scale factor as a function of time, and is given by
\be
a(t)=C_1\sin\left(\frac{\sqrt{3}}{2}mt\right)+C_2\cos\left(\frac{\sqrt{3}}{2}mt\right)\;,
\ee
which has oscillatory behaviour. Note that the solution above has, as expected, three constants due to the fact that $\dot{q}=h(q)$ is a {\it third} order differential equation in $a$.

\subsection{Reconstruction from the dynamics of higher order parameters}

Since the $f(R)$ theories are of order four it is natural to give a relation for  the scale factor involving a fourth order equation for $a$. This would involve naturally  the {\it jolt} parameter $j=\dddot{a}a^{2}/\dot{a}^3$ ~\cite{Dunajski:2008tg}. It  turns out that with this parameter makes it difficult to achieve the results of the 
previous section.

For this reason, in analogy with what is usually done with the scale factor,  we propose the following expansion of the Hubble parameter:
\begin{equation} \label{}
H=H_0+\dot{H}(t-t_0)+\frac{1}{2}\ddot{H}(t-t_0)^2+...\;,
\end{equation}
which allows us to define two new cosmic parameters:
\be
Q=\frac{\dot{H}}{H^2}\qquad J=\frac{\ddot{H}H}{\dot{H}^2}\;.
\ee
It is clear that $Q=-(q+1)$ and $J=\frac{j+3 q+2}{(q+1)^2}$, so 
that they are related to the standard parameters $j$ and $q$.

We can easily check from the above definitions that
\be
\frac{1}{H}\frac{\dot{Q}}{Q^2}=J-2\;.
\ee
Let us now suppose that we would like to reconstruct from the given condition 
\be
\frac{1}{H}\dot{J}=h(J)\;.
\ee
We then integrate the above conditions to find
\be
\frac{1}{Q(J)}=\int\frac{2-J}{h(J)}dJ\;;\;H(J)=\exp\left[\int\frac{Q(J)}{h(J)}dJ\right]\;,
\ee
\be
a(J)=\exp\left[\int\frac{dJ}{h(J)}\right]
\ee
The Ricci scalar can now be written in terms of these quantities as 
\be
R(J)=6H(J)^2(2+Q(J))\;.
\ee
If the above relation is explicitly invertible, that is we have $J=g(R)$, then we can 
again substitute everything into the Friedmann equation to give
\bea
3H[g(R)] R_{,J}[g(R)]H[g(R)]h[g(R)]f''& \nonumber\\
+f'\left\{3H(g(R))^2-\frac{Rf'}{2}\right\}+\frac{f}{2}-\frac{\rho_0}{a[g(R)]^{3(1+w)}}&=0\;.
\eea
Solving this equation gives the required theory.

As a specific example, let us assume the condition to be 
\be
\frac{1}{H}\dot{J}=J-2\;.
\ee
Also for simplicity let us consider a universe filled with dust-like matter.
Solving the Friedmann equation we get the following particular solution:
\be
f(R)=\alpha_1R+\alpha_2R^2+\alpha_3R^3+\alpha_4\;,
\ee
where  $\alpha_n(n=1..4)$ are constants depending on $\rho_0$. Now solving for 
$a(t)$, we obtain the following solution in integral form:
\be
t=\int_0^{a(t)}\frac{dx}{\sqrt{x^4+2x^3+C1}}\;.
\ee 
It is clear that this solution, although given implicitly, contains four different modes.

\section{Discussion and Conclusion}
In this paper we have presented a new approach to reconstruction methods for $f(R)$-gravity. Using the standard and some newly defined cosmic parameters, we have been able to connect  some non-trivial  cosmic histories to some relatively simple $f(R)$ functions,  solutions which would have been harder to obtain through a direct integration of the field equations.  Given the simplicity of the functions $f(R)$ obtained, one can proceed to examine other features of these solutions such as the evolution of cosmological perturbations in these models. 

The few examples presented here are an indication of the richness of the general behavior of $f(R)$-cosmology, which so far could be grasped only via the dynamical system approach. We are confident that this new method will help uncover more details of the cosmology of these theories and lead to a deeper understanding of their features and structure.

\acknowledgments
The authors would like to thank the National Research Foundation (South Africa) for financial support. SC was funded by Generalitat de Catalunya through the Beatriu de Pin\'{o}s contract 2007BP-B1 00136.

\end{document}